\documentclass[runningheads]{svmult}

\usepackage{makeidx}   
\usepackage{graphicx}  
\usepackage{subeqnar}  
\usepackage{multicol}  
\usepackage{physmubb}  
\makeindex	       

\newcommand{\thub}{\tilde t}
\newcommand{\Uhub}{\tilde U}
\newcommand{\Vhub}{\tilde V}
\newcommand{\pmin}{$p_{min}$}
\begin{document}
\title*{Electron Dynamics in AC-Driven Quantum Dots}
\toctitle{Electron Dynamics in AC-Driven Quantum Dots}
%
%
\titlerunning{Electron Dynamics in AC-Driven Quantum Dots}
%
\author{C.E. Creffield$^{1,2}$
\and G. Platero$^1$}
%
\authorrunning{C.E. Creffield 
and G. Platero}
%
%
\institute{1. Instituto de Ciencia de Materiales de Madrid (CSIC), 
Cantoblanco, E-28049, Madrid, Spain \\
2 Dipartimento di Fisica, Universit\`a di Roma
``La Sapienza'', Piazzale Aldo Moro 2, I-00185 Roma, Italy}

\maketitle

\begin{abstract}
We investigate the dynamics of interacting electrons confined
to two types of quantum dot system, when driven by an
external AC field. We first consider a system of two electrons
confined to a pair of coupled quantum dots \cite{cecgp1}
by using an effective two-site model of Hubbard-type.
Numerically integrating the Schr\"odinger equation in time
reveals that for certain values of the
strength and frequency of the field the tunneling
between the dots can be destroyed, thus
allowing the correlated two-electron states to be manipulated. We
then show how Floquet theory \cite{floq} can be used
to predict the field parameters at which this effect occurs.

We then consider the case of confining the electrons to
a single two-dimensional quantum dot in the limit of low
particle-density. In this system the electrons
form strongly correlated states termed Wigner molecules,
in which the Coulomb interaction causes them to become highly
localised in space. Again using an effective model of Hubbard-type,
we investigate how the AC field can drive the dynamics of the Wigner
states \cite{cecgp2}. As before, we find that the AC field
can be used to control the tunneling between various
charge configurations, and we relate this
to the presence of avoided crossings in the Floquet quasi-energy
spectrum. These results hold out the exciting possibility of using
AC fields to control the time evolution of entangled states in
mesoscopic devices, which has great relevance to the rapidly advancing
field of quantum information processing.
\end{abstract}


\newpage

\section{Introduction}

A quantum dot (QD) is a structure in which electrons
can be confined to small length scales, comparable to their Fermi
wavelength. A set of electrons held in such a structure is conceptually
similar to a set of atomic electrons bound to a nucleus, and for
this reason quantum dots are sometimes termed ``artificial atoms''
\cite{kastner}. Unlike real atoms, the physical properties of quantum
dots can be easily varied, which gives theorists and experimentalists the
opportunity to study novel quantum effects in a
well-controlled system.

To extend the atomic analogy further, we can consider linking
QDs together to form ``artificial molecules''.
By allowing electrons to tunnel between the QDs, the
electronic states on the QDs can hybridize, and form new
states that extend over the whole system \cite{blick}.
The degree of the tunneling
determines the strength of this hybridization. If the tunneling
is weak the electrons remain essentially localised on the QDs
in analogy with ionic bonding states. Conversely, if the tunneling is
strong then the electrons form delocalised states with
a covalent character. Recent transport experiments using
AC potentials \cite{oos} have been performed on double QD systems,
and have indeed revealed the ionic or covalent character of the
electronic states by measurement of the induced photo-current.

Ever since the pioneering work of Anderson \cite{anderson}, it
has been known that random spatial disorder can cause electronic
states to become localised in quantum systems. More recently it has
been found that an AC driving field can produce a similar intriguing
effect termed {\em dynamical localisation}, in which
the tunneling dynamics of a particle can be destroyed.
One of the first systems in which this effect was predicted
is that of a particle moving in a double-well
potential \cite{hanggi_prl}. A physical realization of
this could consist of two coupled QDs containing a single electron
--- the simplest type of artificial molecule possible.
If this system is prepared with the electron occupying
one of the QDs, we can expect it to tunnel across to the
other QD on a time scale set by the Rabi frequency. However,
if an AC field of the correct strength and frequency
is applied to the system, the tunneling is destroyed, and the
particle will remain trapped in the initial well.

Weak time-dependent fields are generally treated as small perturbations,
which produce transitions between the eigenstates of the
unperturbed quantum system. This approach, however, is not applicable to
treat the strong driving fields required to produce dynamical localisation,
and instead the technique of Floquet analysis \cite{floq}, which is
valid in all regimes of driving, has proven
to be extremely effective. In this approach, which
we briefly outline in the next section,
the important quantities to calculate are the {\em quasi-energies},
which play a similar role in driven systems to
the eigenenergies in the undriven case. In particular,
dynamical localisation occurs when two quasi-energies of states
participating in the dynamics approach each other, and become either
degenerate (a crossing) or close to degenerate (an avoided crossing).
Using this formalism, analytic and numerical studies of the
double-well system have shown \cite{shirley,holthaus,hanggi_epl}
that in the limit of high frequencies, quasi-energy crossings occur
when the ratio of the field strength
to the frequency is a root of the Bessel function $J_0$.

Adding a second electron to the coupled QD system, however, introduces
considerable complications. At the low electron densities
typically present in QDs, strong correlations produced by the Coulomb
interaction can significantly influence the electronic structure.
One of the most dramatic consequences of this is the formation
of {\em Wigner molecule} states \cite{kramer}.
Understanding the interplay between electron correlations and
the driving field is, however, extremely desirable, as the ability
to rapidly control electrons using AC fields \cite{cole} has immediate
applications to quantum metrology \cite{tamb_prl} and quantum information
processing. In particular, manipulating entangled electrons on short
timescales is of great importance to the field of quantum
computation \cite{divincenzo}.

We study this problem here by applying the
Floquet formalism to systems of {\em interacting} particles.
The first system we consider is that of two interacting electrons
confined to a pair of coupled QDs. A consequence of the interaction
is that the system only responds strongly to
the field when the field frequency is in resonance with the Coulomb
interaction energy. When this condition is satisfied
we find that, as for the single-particle case, coherent destruction
of tunneling (CDT) can again occur, but that it governed by the
roots of the higher-order Bessel functions. We then go
on to consider a {\em two-dimensional} QD in the Wigner regime, which
may also be described by a lattice model of Hubbard-type \cite{jhj_prb}.
Using the same approach we show that CDT can again occur when
similar conditions are satisfied, and we clarify how an applied AC field
can drive charge redistributions within a strongly correlated QD.
Finally we summarize our results and give some brief conclusions.

\section{Methods and approaches}

\subsection{Introduction to Floquet theory}

We consider a general quantum system driven by a periodic electric
field, described by a time-dependent Hamiltonian which we
can divide in the following way:
\begin{equation}
H(t) = H_{t} + H_I + H_{AC}(t), \quad H_{AC}(t) = H_{AC}(t + n T) .
\label{divide}
\end{equation}
Here $H_{t}$ holds the tunneling terms, $H_I$ holds
the electron-electron interaction terms
and $H_{AC}(t)$ describes the interaction of the system with
the $T$-periodic driving field. The periodicity of the driving
field allows us to use the Floquet theorem to write solutions
of the Schr\"odinger equation as $\psi(t) = \exp[-i \epsilon_j t]
\phi_j(t)$ where $\epsilon_j$
is called the quasi-energy, and $\phi_j(t)$ is a function
with the same period as the driving field, called the Floquet
state. This type of expression is familiar in the
context of solid-state physics, where {\em spatial} periodicity permits
an analogous rewriting of the spatial wavefunction in terms of
quasi-momenta and Bloch states (Bloch's theorem).

The Floquet states provide a complete basis, and thus the
time-evolution of a general state may be written as:
\begin{equation}
| \Psi(t) \rangle= \sum_j \ \left( c_j \mbox{e}^{-i \epsilon_j t} \right)
| \phi_j(t) \rangle ,
\label{floq_exp}
\end{equation}
which is formally analogous to the standard expansion
in the eigenvectors of a time-independent Hamiltonian. Indeed, in the
adiabatic limit, $T = 2 \pi / \omega \rightarrow \infty$, the
quasi-energies evolve to
the eigenenergies, and the Floquet states to the eigenstates.
It is important to note that in this expansion
both the basis vectors (the Floquet states) and the expansion
coefficients explicitly depend on time. The nature of this time-dependence
is very different however, and the superposition of the $T$-periodicity
of the Floquet states with the phase factors arising from
the quasi-energies produces a highly complicated, quasi-periodic
time-dependence in general. As the Floquet states
have the same period as the driving field, they are only able to produce
structure in the time-dependence on short time-scales. Consequently,
the dynamics of the system on time-scales much larger than $T$
is essentially determined by
{\em just} the quasi-energies, and hence evaluating the quasi-energies
provides a simple and direct way of investigating
this behavior. In particular, when two quasi-energies approach
degeneracy the time-scale for tunneling between the states diverges,
producing the phenomenon of CDT.

As we shall see for the specific quantum systems we consider, it
is frequently the case that the total Hamiltonian is invariant
under the generalized parity operation:
$x \rightarrow -x; t \rightarrow t + T/2$. As a result
the Floquet states can also be classified into parity
classes, depending  whether they are odd or even under
this parity operation. Quasi-energies belonging to different
parity classes may cross as an external parameter
(such as the field strength) is varied, but
if they belong to the same class the von Neumann-Wigner
\cite{vonneum} theorem forbids this, and the closest approaches possible
are avoided crossings.
Identifying the presence of crossings and avoided crossings in the
quasi-energy spectrum thus provides a necessary (though not sufficient)
condition for CDT to occur.

\subsection{Perturbation theory for Floquet states \label{method}}

Although the quasi-energies are extremely useful for interpretation of
the time-dependence of a quantum system, they are usually difficult
to calculate and numerical methods must be employed. When the driving
field dominates the dynamics, however, it is possible to use
a form of perturbation theory introduced by Holthaus \cite{holthaus},
in which the time-dependent part of the problem is solved exactly, and
tunneling part of the Hamiltonian, $H_t$, acts as the perturbation.
This was generalized to treat interacting systems in
Refs.\cite{cecgp1,cecgp2}
and was found to be very successful in the high-frequency regime,
where $\hbar \omega$ is the dominant energy-scale. We now give a brief
outline of this method.

The Floquet states and their quasi-energies may be conveniently
obtained from the eigenvalue equation:
\begin{equation}
\left( H(t) - i \hbar \frac{\partial}{\partial t}  \right)
| \phi_j(t) \rangle = \epsilon_j | \phi_j(t)  \rangle
\label{floqeq}
\end{equation}
where we consider the operator
$\left[ H(t) - i \hbar \partial / \partial t \right]$ to operate in
an {\em extended} Hilbert space of $T$-periodic functions \cite{sambe}.
The procedure consists of dividing the Hamiltonian as in Eq.\ref{divide},
and finding the eigensystem of the operator
$\left[ H_{I} + H_{AC}(t) - i \hbar \partial / \partial t \right]$,
while regarding the tunneling Hamiltonian $H_{t}$ as acting
as a perturbation. Standard
Rayleigh-Schr\"odinger perturbation theory can now be
used to evaluate the order-by-order corrections to this result,
requiring only that we define an appropriate inner product for
the extended Hilbert space:
\begin{equation}
\langle \langle \phi_m | \phi_n \rangle \rangle_T = \frac{1}{T}
\int_0^{T} \langle \phi_m(t') | \phi_n(t') \rangle dt' .
\label{inner}
\end{equation}
Here $\langle \cdot | \cdot \rangle$ denotes the usual scalar
product for the spatial component of the wavevectors,
and  $\langle \cdot | \cdot \rangle_T$ is the integration
over the compact time coordinate. We shall show in later
sections how this method can be used to obtain analytical forms
which accurately describe the behavior of the quasi-energies
for the systems we study.

\subsection{Numerical methods \label{num}}

To study the time-evolution of each system, we used
a fourth-order Runge-Kutta method to evolve a given initial
state in time --- typically of the order of fifty periods of
the driving field. Throughout the time-evolution,
physical quantities such as the number occupation of a given site
were measured, and it was ensured
that the unitarity of the wavefunction was accurately preserved.

A number of different methods can be used to numerically
calculate the quasi-energies of a quantum system, and a
detailed description of them is given in Ref.\cite{floq}.
One technique well-suited to our approach
is to evaluate the unitary time-evolution
operator for one period of the driving field $U(t+T,t)$,
and then to diagonalize it.
It may be easily shown that the eigenvectors of this operator are
equal to the Floquet states, and its eigenvalues are related to the
quasi-energies via $\lambda_j = \exp[-i \epsilon_j T]$. This method
is particularly convenient for our purposes, as $U(T,0)$ can be
obtained by propagating the unit matrix in time over one period
of the field, using the same Runge-Kutta method described above.

\section{The driven double quantum-dot}

We consider a simplified model of this system,
in which each QD is replaced by a single site.
Electrons are able to tunnel between the sites, and we include
the effect of interactions by means of a Hubbard-$U$ term:
\begin{equation}
H = \thub \sum_{\sigma} \left( c_{1 \sigma}^{\dagger}
c_{2 \sigma}^{ } + \mbox{H.c.} \right) +
\sum_{i=1}^{2} \left(\Uhub n_{i \uparrow} n_{i \downarrow} +
E_i(t) n_i \right).
\label{hubbard}
\end{equation}
Here $\thub$ is the hopping parameter, and for the remainder of
this work we shall take $\hbar$ equal to one, and measure
all energies in units of $\thub$. $E_i(t)$ is the external electric
potential applied to site $i$. Clearly only the potential
difference, $E_1 - E_2$, is of physical importance, so we may choose
to take the symmetric parametrization:
\begin{equation}
E_1(t) = \frac{E}{2} \cos \omega t, \qquad
E_2(t) = -\frac{E}{2} \cos \omega t.
\label{field}
\end{equation}
The Hilbert space of Hamiltonian (\ref{hubbard}) is six-dimensional,
comprising three singlet states and a three dimensional triplet space.
Measurements on semiconductor QDs have shown that the spin-flip
relaxation time is typically extremely long \cite{fujisawa},
and so we have chosen
not to include any spin-flip terms in the Hamiltonian. Consequently
the singlet and triplet sectors are completely decoupled, and so if
the initial state possesses a definite parity this will be retained
throughout its time-evolution, and only states of the same parity
need to be included in the basis.

\begin{figure}
\begin{center}
\includegraphics[width=.75\textwidth]{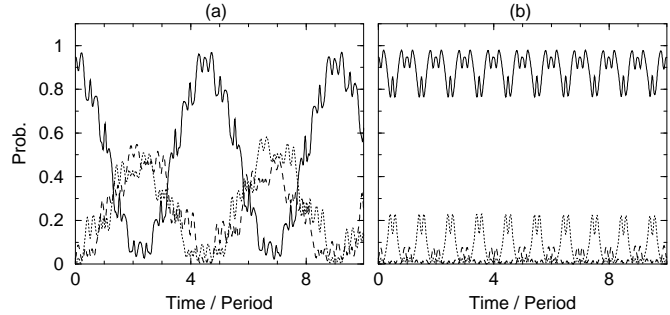}
\end{center}
\caption{Time evolution of the driven double QD system
for $\Uhub=8$ and $\omega=4$.
(a) electric potential, $E = 30.0$; (b) $E = 33.5$.
Thick solid line = $p_{RL}(t)$, dotted line = $p_{LL}(t)$,
dashed line = $p_{RR}(t)$.}
\label{two-site}
\end{figure}

To study the time evolution of the system, we used the ground state
of the static Hamiltonian (a singlet) as the initial state, and evolved it
in time as described in Section \ref{num}. Three probability functions
were measured throughout the time evolution: $p_{LL}(t)$, $p_{RR}(t)$
and $p_{RL}(t)$, which are respectively the probability that both electrons
are in the left QD, both are in the right QD, and that one electron is
in each of the QDs. The Coulomb interaction favors separating the
electrons, and thus for strong interactions the ground-state has a
large value of $p_{RL}$, and relatively small values of $p_{LL}$ and
$p_{RR}$. We show in Fig.\ref{two-site} the time evolution of these
quantities for $\Uhub=8$ and $\omega = 4$, at two different values of
electric potential. In both cases the detailed form of the time-evolution
is highly complicated, but it is clear that the system behaves in two
distinct ways. In Fig.\ref{two-site}a the value of $p_{RL}$
periodically cycles between its initial high value
(indicating that each dot holds approximately one electron)
to nearly zero, while the values of $p_{LL}$ and $p_{RR}$
correspondingly rise and fall at its expense. This behavior is
very different to that shown in Fig.\ref{two-site}b, where $p_{RL}$
never drops below a value of 0.78, and the other two probabilities
oscillate with a very small amplitude. It thus appears that CDT is
occurring in the second case, and that the system's time evolution
is essentially frozen. We shall term the minimum value
of $p_{RL}$ attained during the time-evolution \pmin, and use this
to quantify whether CDT occurs, as a high value of
\pmin signifies that tunneling has been destroyed, while a low value
indicates that the electrons are free to move between the QDs.

\begin{figure}
\begin{center}
\includegraphics[width=.75\textwidth]{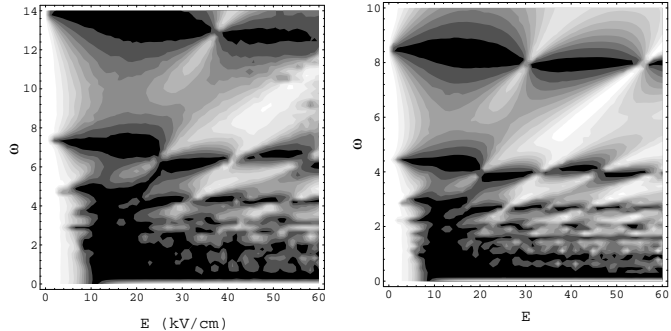}
\end{center}
\caption{\pmin as a function of the strength $E$ and
energy $\hbar \omega$ of the AC field:
(a) for a full simulation of a QD system ($\hbar \omega$ in units of meV)
(b) for the two-site model with $\Uhub=8$ (both axes in units of $\thub$). }
\label{interact}
\end{figure}

\begin{figure}
\begin{center}
\includegraphics[width=.5\textwidth]{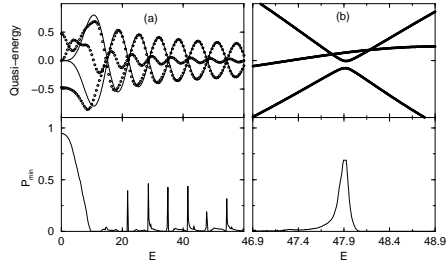}
\end{center}
\caption{(a) Quasi-energy spectrum for the two-site model for
$\Uhub = 8$ and $\omega = 2$, circles = exact results,
lines = perturbation theory,
(b) magnified view of exact results for a single avoided crossing.
Beneath are the corresponding plots of \pmin.}
\label{quasi}
\end{figure}

In Fig.\ref{interact}b we present a contour plot of \pmin
as a function of both of the frequency and strength
of the AC field. Dark areas correspond
to low values of \pmin, and it can be seen that
they form horizontal bands, indicating
that the system is excited strongly by the
AC field only at ``resonant'' values of $\omega$.
Close examination of this plot reveals that these bands occur
at frequencies $\omega = \Uhub, \ \Uhub/2, \ \Uhub/3 \dots$,
at which the system can absorb an integer number of photons
to overcome the Coulomb repulsion between electrons, thereby
enabling tunneling processes such as
$| \uparrow, \downarrow \rangle \ \rightarrow \
| 0, \uparrow \downarrow \rangle$ to occur.
We can additionally observe that these bands are punctuated by
narrow zones in which CDT occurs.
Their form can be seen more clearly in the cross-section of \pmin
given in Fig.\ref{quasi}a, which reveals them to be narrow peaks.
These peaks are approximately equally spaced along each resonance,
the spacing increasing with $\omega$. In Fig.\ref{interact}a we
show another contour plot of \pmin, this time obtained from a
full simulation of detailed physical model of two interacting electrons
confined to a pair of coupled GaAs QDs (for more details on this
simulation see Ref.\cite{cecgp1}). The striking similarity between
these results clearly indicates that our simple, effective model
(\ref{hubbard}) indeed captures the essential processes occurring in
the full system.

We emphasize that these results are radically different to those
obtained for non-interacting particles. In this case an analogous
plot of delocalisation shows a fan-like structure
\cite{hanggi_epl}, in which localisation occurs along lines
given by $\omega = E / x_j$, where $x_j$ is the $j$-th root
of the Bessel function $J_0(x)$. As a test of our method, we repeated
our investigation with the inter-electron Coulomb repulsion
$\Uhub$ set to zero, and found that the fan structure was
indeed reproduced.

\begin{figure}
\begin{center}
\includegraphics[width=.5\textwidth]{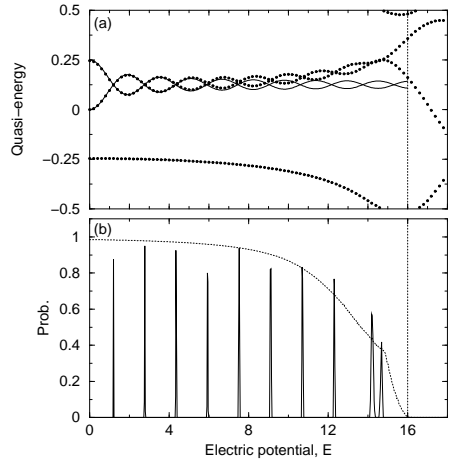}
\end{center}
\caption{(a) Quasi-energy spectrum for the two-site model for
$\Uhub = 16$ and $\omega = 2$, circles = exact results,
solid line = two-level approximation,
$\epsilon_{\pm} = \pm (\Delta/2) J_0(2 E/\omega)$.
(b) Solid line = minimum value of $p_{LL}(t)$, starting
from the initial configuration $|\uparrow \downarrow, 0 \rangle$.
The dotted line denotes \pmin, which shows little structure. In both plots the 
vertical dotted line marks the transition to the strong-driving regime.} 
\label{transition}
\end{figure}

In Fig.\ref{quasi}a we show the Floquet quasi-energies as a function
of the field strength for $\omega = 2$, one of the resonant frequencies
visible in Fig.\ref{interact}b. We see that the system possesses
{\em two} distinct regimes of behavior, depending on whether
the driving potential is weaker of stronger than $\Uhub$.
For weak fields $E < \Uhub$, as studied previously in Ref.\cite{zhang},
the Floquet spectrum consists of one isolated
state (which evolves from the ground state) and two states which
make a set of exact crossings. Although in this regime \pmin
shows little structure, these
crossings do in fact influence the system's dynamics. To demonstrate
this, we show in Fig.\ref{transition} the Floquet
quasi-energies in the weak-field regime for the case of
$\Uhub = 16$, and plot beneath it the minimum value of $p_{LL}$
attained during the time-evolution, where this time the state
$|\uparrow \downarrow, 0 \rangle$ has been used as the initial state.
It can be seen that for this choice of initial condition, the crossings
of the quasi-energies again produce CDT and freeze the initial state
--- despite the Coulomb repulsion between the electrons.

This surprising result may be understood as follows.
For large values of $\Uhub$, the singlet eigenstates of the undriven
system consist of the ground state, separated by the Hubbard
gap $\Uhub$ from two almost degenerate excited states. For
small values of the driving potential, the
two excited states remain isolated from
the ground state, and constitute an effective two-level system with
a level-splitting of $\Delta \simeq 4 \thub^2/\Uhub$. Thus if
the system is prepared in an initial state which projects mainly
onto the excited states, its dynamics will be governed by
the two-level approximation \cite{shirley,holthaus,hanggi_epl},
and CDT will occur at the roots of $J_0$. We show in Fig.\ref{transition}a
the quasi-energies obtained from the two-level approximation, which give
excellent agreement with the actual results with {\em no} adjustable
parameters. As $E$ becomes comparable to the Hubbard gap, however, the two
excited states are no longer isolated from the ground state, and all three
levels must be taken into account. This can be seen in the progressive
deviation of the quasi-energies from the two-level approximation
as the electric potential approaches $\Uhub$.

When the electric potential exceeds $\Uhub$, the system displays
a very different behavior, in which \pmin remains close to zero
except at a series of narrow peaks,
corresponding to the close approaches of two of the quasi-energies.
A detailed examination of these approaches (see Fig.\ref{quasi}b)
reveals them to be {\em avoided crossings} between the Floquet states
which evolve from the ground state and the higher excited state, and
have the same generalized parity. The remaining state, of opposite
parity, makes small oscillations around zero, but its exact crossings
with the other two states do not correlate with any structure in \pmin.

To interpret this behavior in the strong-field regime,
we now obtain analytic expressions for
the quasi-energies via the perturbation theory described in
Section \ref{method}. The first step is to solve the eigenvalue
equation (\ref{floqeq}) in the absence of the tunneling component $H_t$.
In a real-space representation the interaction terms are diagonal,
and so it can be readily shown that an orthonormal
set of eigenvectors is given by:
\begin{eqnarray}
|\epsilon_0(t)\rangle &=& \left( \exp \left[i \epsilon_0 t \right], \ 0, \ 0
\right)
\nonumber \\
|\epsilon_+(t)\rangle &=& \left( 0, \ \exp \left[-i (\Uhub - \epsilon_+) t
+ i \frac{E}{\omega} \sin \omega t \right], \ 0 \right)
\nonumber \\
|\epsilon_-(t)\rangle &=& \left( 0, \ 0, \ \exp \left[-i (\Uhub - \epsilon_-) t
- i \frac{E}{\omega} \sin \omega t \right] \right)
\end{eqnarray}
Imposing $T$-periodic boundary conditions reveals the corresponding
eigenvalues (modulo $\omega$) to be $\epsilon_0 = 0$ and
$\epsilon_{\pm} = \Uhub$.
These eigenvalues represent the zeroth-order approximation to the Floquet
quasi-energies, and for frequencies such that $\Uhub = n \ \omega$
all three eigenvalues are degenerate.
This degeneracy is lifted by the perturbation $H_t$, and
to first-order, the quasi-energies are obtained by diagonalizing
the perturbing operator
$P_{ij} = \langle \langle \epsilon_i | H_t | \epsilon_j \rangle \rangle_T$.
By using the well-known identity:
\begin{equation}
\exp\left[-i \beta \sin \omega t \right] = \sum_{m=-\infty}^{\infty}
J_m (\beta) \exp \left[-i m \omega t \right],
\end{equation}
to rewrite the form of $|\epsilon_{\pm}(t)\rangle$, the matrix elements of
$P$ can be obtained straightforwardly, and its eigenvalues subsequently
found to be $\epsilon_0 = 0$ and $\epsilon_{\pm} = \pm 2 J_n(E/\omega)$.
Fig.\ref{quasi}a demonstrates the excellent agreement
between this result (with $n = 4$) and the exact quasi-energies
for strong and moderate fields, which allows the position of the peaks
in \pmin to be found by locating the roots of $J_n$. Similar excellent
agreement occurs at the other resonances.
For weak fields, however, the interaction terms do not
dominate the tunneling terms and the perturbation theory breaks down,
although we are still able to treat the system phenomenologically
by using the effective two-level approximation.

\section{The square quantum dot}

\subsection{What is a Wigner molecule?}
As we remarked in the Introduction, the Coulomb interaction between
the electrons can significantly affect the electronic structure
of a QD. Such strongly correlated problems are
notoriously difficult to treat, and the addition of a time-dependent
field complicates the problem even further.
When the mean inter-electron separation exceeds a certain critical value,
however, a surprising simplification occurs, as the Coulomb
interaction dominates the kinetic energy and drives
a transition to a quasi-crystalline arrangement which
minimizes the total electrostatic energy. In analogy to the
phenomenon of Wigner crystallization in bulk two-dimensional
systems \cite{wigner,tanatar} such a state is termed a
{\em Wigner molecule} \cite{kramer}.
As the electrons in the Wigner state are sharply
localised in space, the system can be naturally and
efficiently discretized by placing
lattice points just at these spatial locations.
A many-particle basis can then be constructed by taking Slater
determinants of single-particle states defined on these lattice
sites, from which an effective Hamiltonian of Hubbard-type can be generated
to describe the low-energy dynamics of the system \cite{jhj_prb}.
A major advantage of this technique over standard discretization
\cite{akbar,studart} schemes,
in which a very large number of lattice points is taken to approximate
the continuum limit, is that
the dimension of the effective Hamiltonian is much smaller
(typically by many orders of magnitude), which permits the
investigation of systems which would otherwise be prohibitively complex.
This approach has proven to be extremely successful in treating a variety of
static problems, including one-dimensional
QDs \cite{jhj_prb}, two-dimensional QDs with polygonal
boundaries \cite{creff_wig,creff_mag},
and electrons confined to quantum rings \cite{phys_wolf,kosk_ring}.
We further develop this method here by including a
time-dependent electric field, and study the temporal dynamics of
the system as it is driven out of equilibrium.

\subsection{Model and Methods}
\begin{figure}
\begin{center}
\includegraphics[width=.5\textwidth]{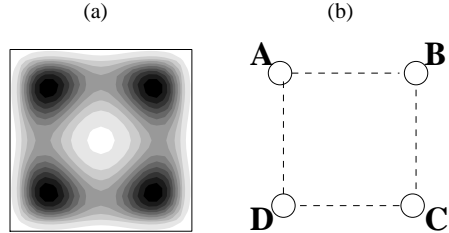}
\end{center}
\caption{(a) Ground-state charge-density for a two-electron square
QD. GaAs material parameters are used, and
the side-length of the QD is 800 nm, placing it in the Wigner regime.
The dark areas indicate peaks in the charge-density.
(b) Lattice points used for the effective lattice-Hamiltonian.}
\label{states}
\end{figure}

We consider a system of two electrons confined to a square QD with a
hard-wall confining potential --- a simple representation of a
two-dimensional semiconductor QD. Such a system can be produced
by gating a two-dimensional electron gas confined at a heterojunction
interface, and by placing a gate split into four quadrants over
the heterostructure \cite{4gates}, the potentials at the corners
of the QD can be individually regulated.
In Fig.\ref{states}a we show the ground-state charge-density
obtained from the exact diagonalization of a square QD \cite{creff_wig},
for device parameters placing it
deep in the Wigner molecule regime. It can be seen that the charge-density
is sharply peaked at four points, located close to the vertices of the QD.
This structure arises from the Coulomb interaction between the electrons,
which tends to force them apart into diagonally opposite corners of the dot.
As there are two such diagonal states, degenerate in energy,
we can understand the form of the ground-state by considering it to be
essentially a superposition of these two states (with a small admixture of
higher energy states). The four points at which the peaks occur
define the sites on which the effective lattice-Hamiltonian operates,
as shown in Fig.\ref{states}b.

We take an effective lattice-Hamiltonian of the form:
\begin{equation}
H =  \sum_{\langle i, j \rangle, \sigma} \left[ \thub \ \left(
c_{i \sigma}^{\dagger} c_{j \sigma}^{ } + \mbox{H.c.} \right) +
\Vhub \ n_i n_j \right]
+ \sum_{i} \left[ \Uhub \ n_{i \uparrow} n_{i \downarrow} +
E_i(t) n_i \right] .
\label{hamiltonian}
\end{equation}
Here $\Vhub$ represents the Coulomb repulsion between electrons
occupying neighboring sites, and $\Uhub$ is the
standard Hubbard-$U$ term, giving the energy cost for double-occupation of a
site. As before, $E_i(t)$ denotes the electric potential at site $i$,
which in general can have a static and a time-dependent component. In
experiment, static offsets can arise either from deviations of the confining
potential of the QD from the ideal geometry, or by the application
of gating voltages to the corners of the QD. Applying corner potentials in
this way would substantially enhance the stability of the Wigner molecule
state, and could also be used to ensure that the multiplet of states included
in the effective lattice-model is well-separated from the other excited
states of the QD system. In this work, however, we do not explicitly consider
the effects of static gates, and we neglect the influence of small,
accidental offsets encountered in experiment as we expect them to have only
minor effects, and indeed may even stabilize CDT \cite{stockberger}. For
convenience, we consider applying an AC field aligned with the $x$-axis of
the QD, which can be parameterized as:
\begin{equation}
E_A = E_D  = \frac{E}{2} \cos \omega t, \qquad
E_B = E_C = - \frac{E}{2} \cos \omega t
\end{equation}
where A,B,C,D label the sites as shown in Fig.\ref{states}b.
We emphasize that although we have the specific system of
a semiconductor QD in mind,
the effective-Hamiltonian we are using can describe a wide range of
physical systems, including $2 \times 2$ arrays of connected
QDs \cite{kotlyar}, and our results are thus of general applicability.

\begin{figure}
\begin{center}
\includegraphics[width=.5\textwidth]{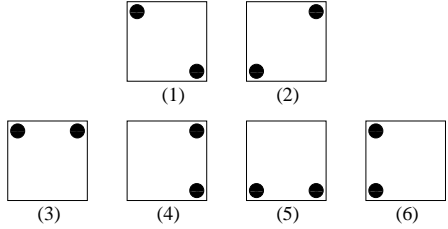}
\end{center}
\caption{Schematic representation of the
two-particle basis states for the singlet sector of the
Hamiltonian. The ground state of the QD is
approximately a superposition of states (1) and (2).}
\label{basis}
\end{figure}

As for the case of the double QD, we include no
spin-flip terms in (\ref{hamiltonian}) and so the singlet and
triplet sectors are again decoupled.
We choose to use initial states with singlet symmetry,
which corresponds to the symmetry of the system's ground-state.
Simple state counting reveals that the singlet sector
has a dimension of ten, and can be spanned by the six states
shown schematically in Fig.\ref{basis}, together with the four
states in which each site is doubly-occupied.

\section{Results}

\subsection{Interacting electrons, double occupancy excluded}
\label{sec_Uinf}

\begin{figure}
\begin{center}
\includegraphics[width=.5\textwidth]{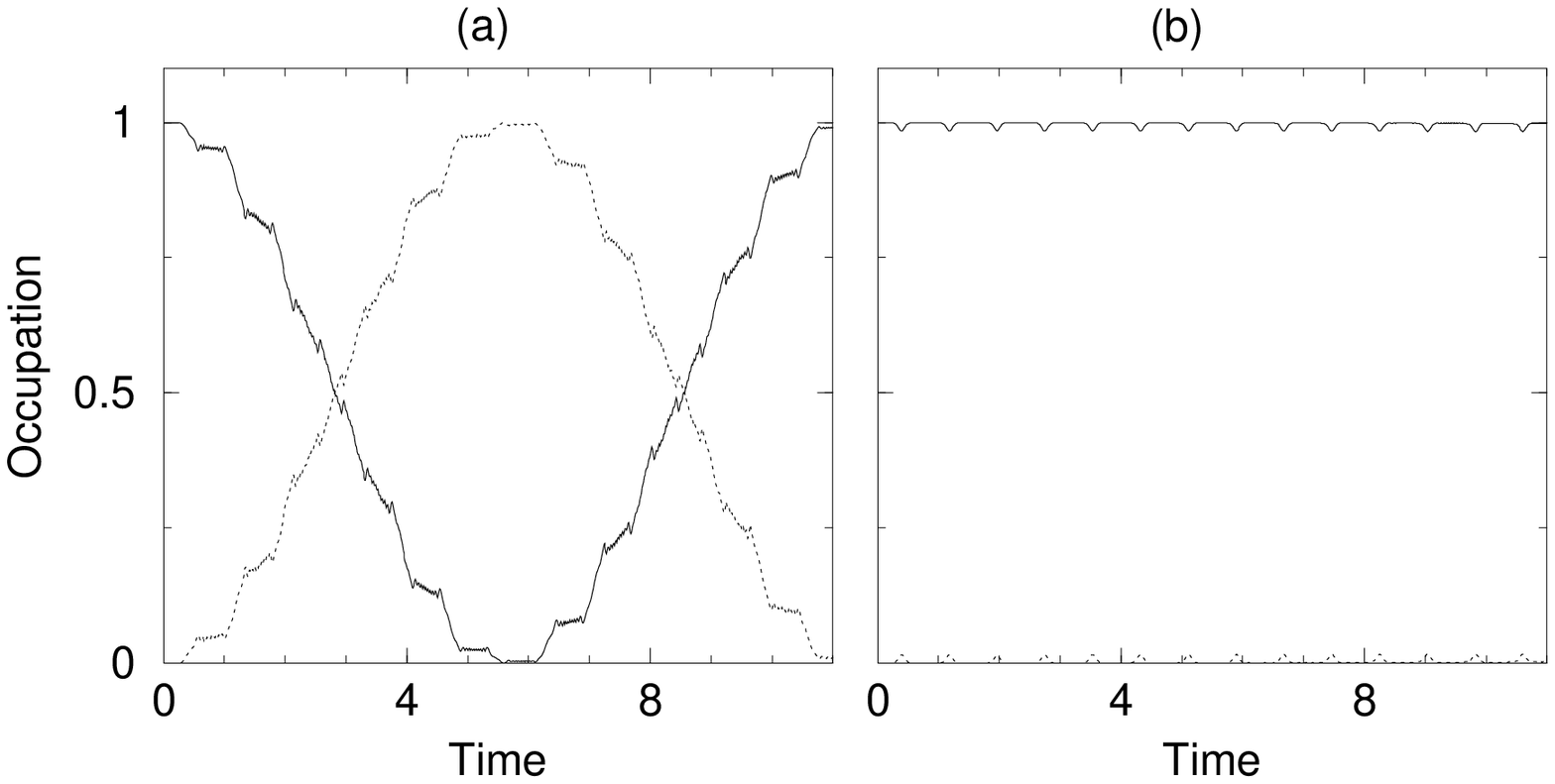}
\end{center}
\caption{Time development of the system for $\Uhub$ infinite,
$\Vhub=80$ and $\omega=8$:
(a) electric potential, E = 100.0
(b) E = 115.7.
Solid line indicates the occupation of sites A and D, the dotted line
the occupation of sites B and C.}
\label{Uinf}
\end{figure}

We begin our investigation by taking the Hubbard-$U$ term
to be infinitely large --- that is, to work in the
sub-space of states with no double occupation. Our Hilbert
space is thus six-dimensional, and we use the
states shown in Fig.\ref{basis} as a basis. We show in
Fig.\ref{Uinf} the time-dependent number
occupation of the four sites at two different values of $E$,
in both cases using state $(6)$ as the initial
state, and setting the AC frequency to $\omega = 8$. In Fig.\ref{Uinf}a
$E$ has a value of 100.0, and it can be clearly seen that the electrons
perform driven Rabi oscillations between the left side of the
QD and the right. Accordingly, the occupation number of the sites
varies continuously between zero and one.
In Fig.\ref{Uinf}b, however, we see that changing the electric
potential to a value of $E=115.7$ produces dramatically different behavior.
The occupations of sites A and D only vary
slightly from unity, while sites B and C remain essentially empty
throughout the time-evolution. Only a small amount of charge can transfer
per period of the driving field between the left and right sides
of the system, producing the small spikes
visible in this figure. The amplitude
of these features is extremely small, however, indicating that
the tunneling between left and right sides has been almost
totally destroyed.

\begin{figure}
\begin{center}
\includegraphics[width=.5\textwidth]{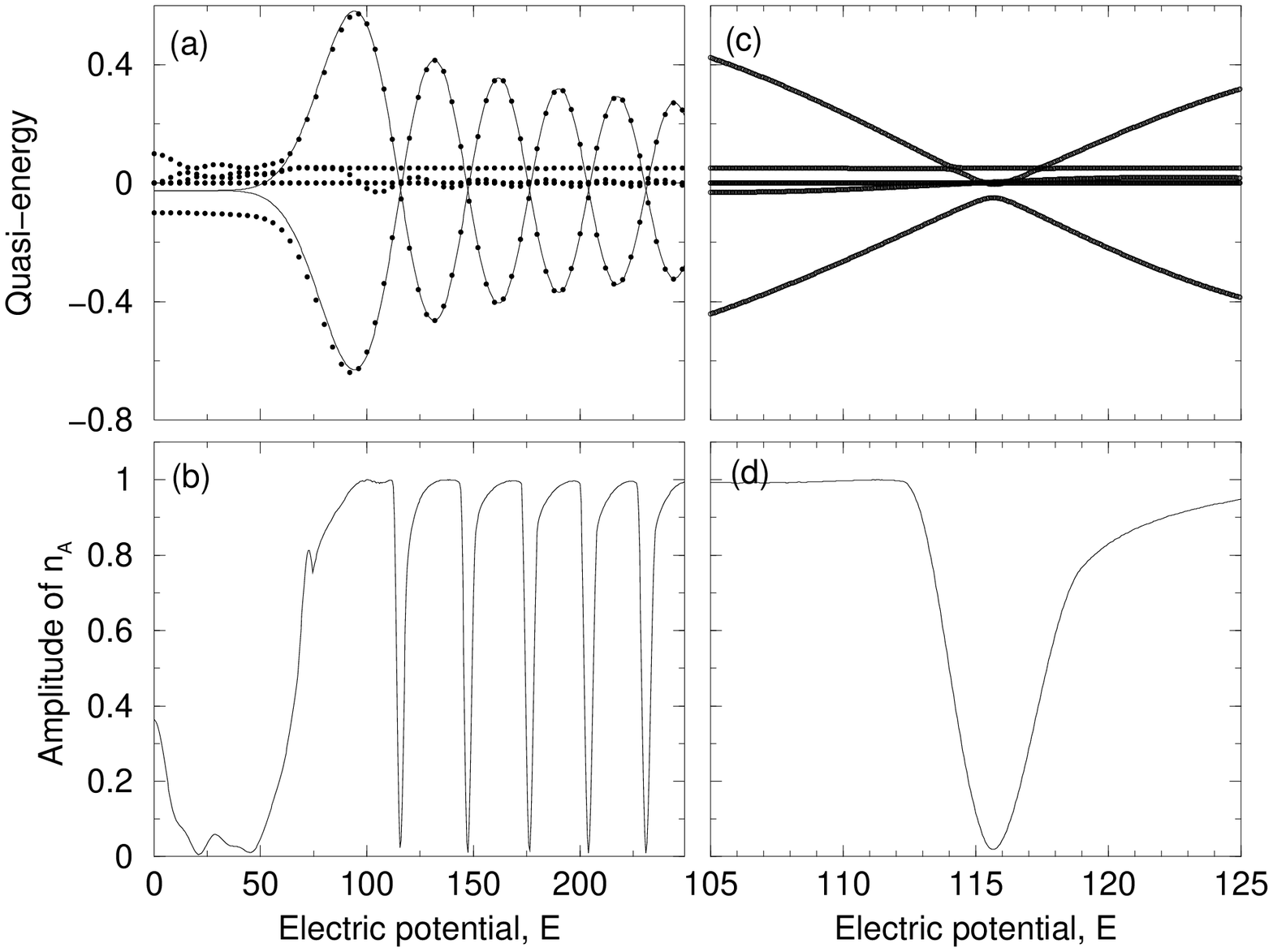}
\end{center}
\caption{(a) Quasi-energies of the system for $\Uhub$ infinite,
$\Vhub=80$ and $\omega=8$: circles = exact results,
lines=perturbative solution [$\pm 2 J_{10}(E/\omega)$].
(b) Amplitude of oscillation of the occupation of site A.
(c) Detail of quasi-energy spectrum, showing an avoided crossing.
(d) Detail of amplitude of oscillations.}
\label{Uinf_floq}
\end{figure}

To confirm that CDT is occurring, we present in Fig.\ref{Uinf_floq}
a comparison of the amplitude of the oscillations of $n_A$ with the
quasi-energy spectrum, as a function of the electric potential $E$.
Similarly to the double QD system, we can see in Fig.\ref{Uinf_floq}a
that the quasi-energies have two different regimes of behavior.
The first of these is the weak-field regime,
$E < \Vhub$, at which the driving field does not dominate the dynamics.
In this regime the quasi-energy spectrum, and correspondingly, the amplitude
of oscillations shows little structure.
The second regime occurs at strong values of potential,
$E > \Vhub$, for which the quasi-energy
spectrum clearly shows a sequence of close approaches.
In Fig.\ref{Uinf_floq}c we show an enlargement of one of these
approaches which reveals it to be an {\em avoided crossing}.
Employing the perturbative method described in Section \ref{method}
demonstrates that the two quasi-energies
involved in these avoided crossings are described by $\pm 2 J_n(E/\omega)$,
where $n$ is equal to $\Vhub / \omega$. We may thus again think of $n$ as
signifying the number of photons the system needs to absorb to overcome
the Coulomb repulsion between the electrons occupying neighboring sites.
The results in Fig.\ref{Uinf_floq}b and Fig.\ref{Uinf_floq}d
clearly show that the locations of the avoided crossings
correspond exactly to quenching of the oscillations in $n_A$,
and so confirm that CDT indeed occurs at these points.

\subsection{Interacting electrons, double-occupancy permitted}

We now take the most general case, and consider the competition between
the $\Uhub$ and $\Vhub$ terms. Setting $\Uhub$ to a finite value
means that the four doubly-occupied basis states are no longer
energetically excluded from the dynamics, and accordingly
we must take the full ten-dimensional basis set.

\begin{figure}
\begin{center}
\includegraphics[width=.5\textwidth]{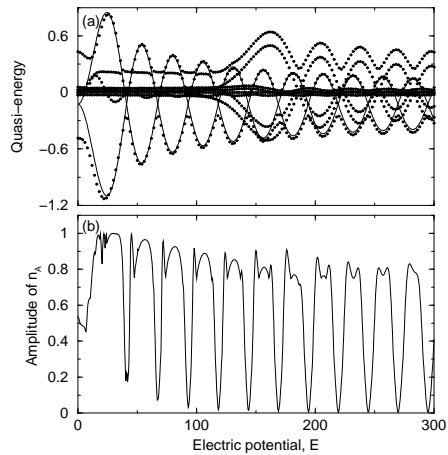}
\end{center}
\caption{(a) Quasi-energies of the system for  $\Uhub=160$ and
$\Vhub=16$, $\omega=8$:
circles = exact results, lines=perturbative solution
[$\pm 2 J_{2}(E/\omega)$].
(b) Amplitude of oscillation of the occupation of site A,
with (6) as the initial state.}
\label{U160_1}
\end{figure}

Although it is difficult to obtain precise estimates for the values of
parameters of the effective Hamiltonian, it is clear that
in general $\Uhub > \Vhub$. Accordingly we choose the
parameters $\Uhub = 160, \ \Vhub=16$ to separate
the two energy-scales widely for our investigation.
We again set the frequency of the AC field to
$\omega=8$, and in Fig.\ref{U160_1}a we show the quasi-energy
spectrum obtained by sweeping over the field strength. It is immediately
clear from this figure that for electric potentials $E < \Uhub$ the
form of the spectrum is extremely similar to the infinite-$\Uhub$
case. Performing perturbation theory confirms that,
as in the previous case, the behavior of the quasi-energies is
given by $\pm 2 J_n(E/\omega)$ where $n=\Vhub / \omega$.
We show in Fig.\ref{U160_1}b the amplitude
of the oscillations of $n_A$ when the system is
initialized in state $(6)$, which demonstrates that at the locations of the
avoided crossings the tunneling parallel to the field is again quenched.

When the electric potential exceeds the value of $\Uhub$, however,
new structure appears in the quasi-energy spectrum.
A group of four quasi-energies, that for weaker fields cluster around
zero, become ``excited'' and make a sequence of avoided crossings
as the field strength is increased.
Perturbation theory predicts that these high-field
quasi-energies are given by $\pm 2 J_m(E/\omega)$, where
$m=(\Uhub-\Vhub)/\omega$, and thus these
avoided crossings arise when the absorption of $m$
photons equates to the electrostatic energy
difference between the two electrons being on neighboring
sites, and doubly-occupying one site. This then indicates
that this structure arises from the coupling of the AC field
to the doubly-occupied states.

\begin{figure}
\begin{center}
\includegraphics[width=.5\textwidth]{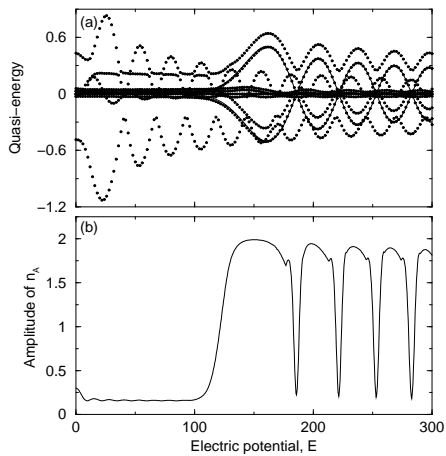}
\end{center}
\caption{(a) Quasi-energies of the system for  $\Uhub=160$ and
$\Vhub=16$, $\omega=8$: circles = exact results,
lines=perturbative solution [$\pm 2 J_{18}(E/\omega)$].
(b) Amplitude of oscillation of the occupation of site A,
with site A doubly-occupied as the initial state.}
\label{U160_2}
\end{figure}

To probe this phenomenon, we time-evolve the system from
an initial state consisting of {\em two} electrons occupying site A.
In Fig.\ref{U160_2}b it can be seen that for electric potentials
weaker than $\Uhub$ the amplitude of the oscillations
in $n_A$ remains small, and shows little dependence on the
field. As the potential exceeds $\Uhub$, this picture changes, and
the AC field drives large oscillations in $n_A$, and in fact mainly
forces charge to oscillate between sites A and B. At the
high-field avoided crossings, however, the tunneling between
A and B is suppressed, which shuts down this process. Instead,
the only time-evolution that the system can perform consists
of {\em undriven} Rabi oscillations between sites A and D,
perpendicular to the field. As these oscillations are undriven they
have a much longer time-scale than the forced dynamics, and thus during
the interval over which we evolve the system
the occupation of A only changes by a small amount, producing the
very sharp minima visible in Fig.\ref{U160_2}b, centered on the roots of
$J_m(E/\omega)$.

\begin{figure}
\begin{center}
\includegraphics[width=.75\textwidth]{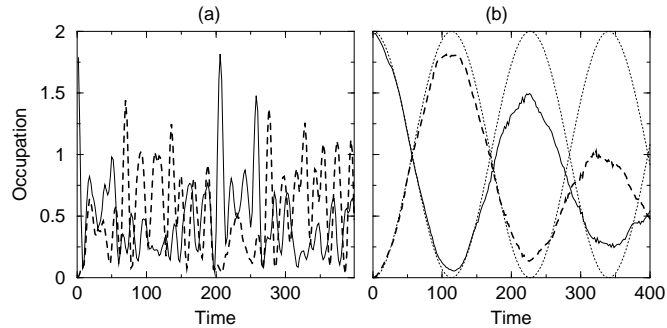}
\end{center}
\caption{Time development of the $\Uhub=160$
and $\Vhub=16$ system for $\omega=8$:
(a) electric potential, E = 200.0 (no CDT)
(b) E = 185.8 (CDT).
Thick solid line indicates the occupation of site A, the
thick dotted line the occupation of site D.
Dotted lines in (b) show the Rabi oscillations of the
isolated two-site system, Eq.\ref{rabi}.}
\label{decohere}
\end{figure}

As the tunneling perpendicular to the field is undriven, it is
straightforward to evaluate the time evolution of the initial state,
if we assume that the left side of the QD is
completely decoupled from the right side. The occupation
of sites A and D is then given by:
\begin{equation}
n_A(t) = 1 + \cos \Omega_R t, \quad n_D = 1 - \cos \Omega_R t
\label{rabi}
\end{equation}
where $\Omega_R = 4 \thub^2/(\Uhub - \Vhub)$.
In Fig.\ref{decohere} we display the occupations of sites A and D as a
function of time, for two values of electric potential. At the first value,
$E=200$, tunneling between the left and right sides of the QD is not
quenched, and accordingly the occupation of the two sites varies rapidly
between zero and two as the electrons are driven by the AC field around
the system. The second value, $E=185.8$, corresponds to the first
high-field avoided crossing. It can be clearly seen that the charge
oscillates between sites A and B, with a frequency of $\Omega_R$.
These Rabi oscillations are damped, however, indicating that the isolation
between the left and right sides of the QD is not perfect.
In this sense we can regard the two sites $B$ and $C$ as providing
an environment, causing the quantum system composed of sites
A and D to slowly decohere in time. When the tunneling between the left
and right sides of the QD is strong, for example at $E=200$, this
decoherence occurs very rapidly. By moving
to an avoided crossing, however, and suppressing the tunneling,
the rate of mixing between the two sides of the QD can be considerably
reduced, and is just limited by the separation in energy between
the two quasi-energies. Tuning the parameters of the
driving field therefore gives us a simple and
controllable way to investigate how a two-electron
wavefunction can decohere in a QD.

\section{Conclusions}

In this work we have studied the interplay between Coulomb interactions
and an AC driving field in two different configurations
of QDs. For the case of the double QD system
we have found that the system presents two different
types of behavior, depending on whether the driving field dominates
the interaction energy. For weak driving, we have found the
surprising result that a suitable AC field can nonetheless freeze the
time-evolution of a doubly-occupied QD {\em despite} the Coulomb interaction,
and that this can be understood by means of an effective two-level model.
At high field strengths the system is only driven strongly
by the field at frequencies for which an integer number of quanta, $n$,
is equal to the interaction energy. When this condition is satisfied
CDT is again able to occur at certain well-defined parameters
of the field, and by using Floquet theory we have shown that these
points correspond to the roots of $J_n(E/\omega)$.

Strong electronic correlations allowed us to use an effective lattice
model of just four sites to treat the square QD, by taking
advantage of the natural discretization of the system in a Wigner molecule
state. In the effective model, the inter-electron Coulomb interaction
is described by two parameters, $\Uhub$ and $\Vhub$, and the
dynamics of the system consists essentially of tunneling from corner to
corner, along the perimeter of the QD. We find again
that when the frequency of the driving field
is in resonance with the Coulomb gap (that is, $m \omega = \Vhub$ or
$m \omega = (\Uhub - \Vhub)$) charge is able to circulate freely around the
system, except at sharply defined field strengths at which tunneling
parallel to the field is destroyed. Floquet theory again proved an excellent
tool to understand this behavior, and revealed
that these points correspond to the roots of $J_m(E/\omega)$.

We have thus shown that AC-fields may not only be used as a spectroscopic
tools to probe the electronic structure of QD systems, but can also be
used to dynamically control the time-evolution of the system. Possible
applications of CDT range from stabilising the leakage of trapped
electrons in physical realisations of quantum bits, to acting as ``electron
tweezers'' by destroying or restoring tunneling between regions of
a mesoscopic device. The tunability of the CDT
effect, and its ability to discriminate between
doubly-occupied and singly-occupied states, make it an
excellent means for rapid manipulation of the
dynamics of strongly correlated electrons in mesoscopic systems.

\section{Acknowledgments}

This work was supported by the Spanish DGES grant MAT2002-02465, by the
European Union TMR contract FMRX-CT98-0180 and by the European
Community's Human Potential Programme under
contract HPRN-CT-2000-00144, Nanoscale Dynamics.\\

%

\end{document}